\documentclass[onecolumn, 12pt]{article}
\usepackage{graphicx}
\usepackage{amsmath}
\usepackage[hidelinks]{hyperref}
\usepackage[margin=0.75in]{geometry}
\usepackage{float} % in the preamble
\usepackage{siunitx}
\usepackage{booktabs}
\usepackage{amssymb}
\usepackage{caption}

\DeclareUnicodeCharacter{2212}{-}
\DeclareUnicodeCharacter{2009}{\,}


\begin{document}

\title{Imaging atomic scale stochasticity: noise in the nearly commensurate charge density wave of 1T-TaS$_2$ for leaky-integrator-and-fire neuromorphic emulator}

\author{Michael Verhage$^1$\\ \parbox{\linewidth}{\small$^1$Department of Applied Physics and Science Education, Molecular Materials to Optoelectronic Devices, Eindhoven University of Technology, The Netherlands}}

\date{}
\maketitle

\begin{abstract}
Low-dimensional materials are promising materials for emulating neuron-like firing were understanding of the atomic-scale origin of this behavior can benefit integration into neurormophic hardware. Using scanning tunneling microscopy (STM) as a highly local probe of correlations in the charge density wave (CDW), we study atomic-scale CDW noise in $1T\text{-}\mathrm{TaS}_2$ at 
room temperature. We statistically analyze the tunnel-junction noise and, to explain the temporally correlated current bursts, introduce a mesoscopic model in which the fractal structure and reorganization of CDW discommensurations are related. This noise mimics leaky-integrate-and-fire neuron dynamics. Our results show that atomic-scale current bursts in the tunnel junction can act as effective nonlinear elements in spiking neural emulators, enabling bioinspired processing through rich CDW dynamics at room temperature and supporting the miniaturization of spiking devices.
\end{abstract}

%++++++++++++++++++++++++++++++++++++++++++++++++++++++++++++
\section{Introduction}
%++++++++++++++++++++++++++++++++++++++++++++++++++++++++++++

The pursuit of biomimetic, energy-efficient hardware has positioned two-dimensional (2D) materials as leading candidates for next-generation neuromorphic devices \cite{Wang2019ArtificialApplication, Sangwan2015GatetunableMoS2, Sangwan2018MultiterminalDisulfide}.
Their atomic thickness and the transport tunability of van der Waals heterostructures enable core neural primitives such as leaky integrate-and-fire (LIF) dynamics \cite{Hao2020MonolayerNetworks} and defect-mediated synaptic plasticity \cite{Wang2026HomogeneousVision}. Recent 2D devices already reproduce dendritic emulation \cite{Oh2025HighlyClassification} and short- and long-term memory \cite{Singh2024BraininspiredNetworks?}, bridging simple electronic switches with the rich temporal dynamics of biological networks \cite{Ko2025TwodimensionalTechnology}. A particularly attractive route exploits the non-linear transport response of strongly correlated 2D materials, in which the electronic order itself supplies the required dynamics \cite{Ye20262DComputing, Liu2021TantalumProperties, Mohammadzadeh2021RoomDevices, Khitun2017TwoDimensionalDevices}.

Among strongly correlated 2D systems, the quasi-2D layered dichalcogenide $1T$-$\text{TaS}_2$ is a compelling candidate, combining charge-density-wave (CDW) order with a rich phase diagram~\cite{Sipos2008Mott1T-TaS2}. This order supports memristive switching~\cite{Yoshida2015MemristiveCrystals, Brown2026ElectricallyDevices}, including bio-relevant optically driven switching~\cite{Zhu2018LightTunableOscillators}, and can serve as a building block for neuromorphic devices \cite{Xiao2024RecentChips, Khitun2017TwoDimensionalDevices}. At \SI{300}{\kelvin}, its nearly commensurate CDW (NC-CDW) phase separates into strongly correlated insulating domains partitioned by a contiguous network of metallic domain walls, or discommensurations \cite{Cho2016Nanoscale1T-TaS2, Cho2017Correlated1T-TaS2}. This provides a directly accessible, electrically reconfigurable built-in network of conducting channels \cite{Brown2026ElectricallyDevices}, and can function as an interesting substrate to emulate neuromorphic behavior.

Device-level studies have established the neuromorphic value of the NC-CDW
phase \cite{Khitun2017TwoDimensionalDevices}. Voltage-tunable oscillators that emulate neuronal spiking exploit the hysteretic NC-to-incommensurate (NC-to-IC) transition \cite{Liu2016ChargedensitywaveTemperature}, and their cycle-to-cycle variability has been mapped onto biological firing regimes, including the stochastic LIF patterns of the mammalian superior olivary complex \cite{Liu2019EnergyEfficientComputing}. These studies, however, probe macroscopic transport, in-plane or cross-plane currents that average over the NC-CDW structure and attribute the nonlinear response to collective CDW depinning \cite{Liu2018LowFrequencyMaterials, Mohammadzadeh2021RoomDevices} or to NC-to-IC phase transitions \cite{Geremew2020HighfrequencyConcept, Brown2026ElectricallyDevices}. This exposes a key problem: the atomic-scale CDW dynamics underlying these nonlinear current fluctuations and thus the neuromorphic functionality cannot be resolved from macroscopic transport alone. Scanning tunneling spectroscopy (STS) offers local access to the CDW density of states \cite{Fei2022Understanding1T-TaSe2}, but at room temperature thermal broadening ($\sim 3k_\mathrm{B}T$) washes out the domain and domain-wall LDOS contrast \cite{Park2019EmergentWave}. In this work we show, while this broadening erases the time-averaged spectroscopic fingerprint, it does not erase the temporal statistics of the tunneling current, which is the observable we exploit here.

In this work, we use tunneling current fluctuations as a local transducer of the CDW phase-order dynamics. This phase order encodes the (in)commensurate structure of the CDW field, $\psi(\mathbf{r}) = \sum_{j=1}^{3} \Delta_j(\mathbf{r})\,
    e^{i\left(\mathbf{Q}_j \cdot \mathbf{r} + \phi_j(\mathbf{r})\right)}$ where $\Delta_j(\mathbf{r})$ is the spatially dependent amplitude, $\mathbf{Q}_j$ the nesting wavevectors, and $\phi_j(\mathbf{r})$ the local
phase field \cite{McMillan1976TheoryTransition}. The three coupled phase fields shift under strain, electric field, or doping, and the discommensuration marks the transition between commensurate domains and the incommensurate matrix. Scanning tunneling microscopy (STM) grants direct access to both this phase order \cite{Verhage2026Interplay1T-TaS2, Pasztor2019HolographicParameter} and its long-timescale dynamics, allowing us to probe current fluctuations directly within the NC-CDW state.

We show that noise-power spectroscopy exhibits a pronounced peak maximum near $200$~mV in the empty states. Bias-dependent noise-power spatial imaging shows that domain have strong contrast at $200$~mV but this contrast collapses at $25$~mV. Therefore we identify the fingerprint of NC-CDW electronic phase separation at room temperature, without needing tunneling spectroscopy \cite{Park2019EmergentWave}. We note that the ground state of $1T$-$\text{TaS}_2$ remains debated, the $\sim 0.2$~eV empty-state feature having been attributed either to the UHB of a Mott state or to a stacking- and termination-dependent band edge \cite{Petocchi2022Mott2}. Our central claim, however, does not hinge on this distinction. To explain this bias selectivity, we introduce a mesoscopic two-channel framework linking the fractal discommensuration-network topology~\cite{Verhage2026Interplay1T-TaS2} to bias-dependent tunneling noise. The first, structural channel treats domain walls as a one-dimensional elastic manifold driven through a two-dimensional quenched-disorder landscape \cite{Park2021Zoology1T-TaS2} (e.g., sulfur or strain vacancies \cite{Verhage2026Interplay1T-TaS2} or thermal-quench disorder \cite{DeLaTorre2025DynamicQuench}), which reorganizes in discrete avalanches. The second, readout channel explains why these avalanches become visible only at the UHB: resonant tunneling into a narrow, spatially confined correlation state is far more sensitive to gap and domain reconfiguration than tunneling into the broad metallic-wall continuum. The UHB resonance thus acts as a nonlinear amplifier, converting rearrangements of the jammed fractal network into the observed current bursts, while the same avalanches leave the wall dominated $25$~mV channel comparatively Gaussian.

Finally, we demonstrate that this CDW noise functions as an effective emulator for a leaky integrate-and-fire (LIF) neuron, a circuit model in which charge accumulates on a leaky membrane until it reaches a threshold that induces a spike and subsequent reset \cite{Hao2020MonolayerNetworks}. By combining numerical LIF simulations with an analog $RC$-circuit hardware emulator \cite{Khitun2017TwoDimensionalDevices}, we connect the underlying materials-level physics of these fluctuations to their neuromorphic behavior. Because leptokurtic input generates infrequent but large threshold-crossing events, the domain-localized bursts operate as a structured entropy source that, unlike Gaussian or memoryless Poisson noise, inherently yields sparse, irregular spike trains with highly variable inter-spike intervals. The tip bias thus serves as a single tuning parameter that adjusts the firing pattern from regular (wall, Gaussian) to burst-like (domain, leptokurtic), providing a pathway toward bias-tunable neuromorphic hardware.

%++++++++++++++++++++++++++++++++++++++++++++++++++++++++++++
\section{Results and Discussion}
%++++++++++++++++++++++++++++++++++++++++++++++++++++++++++++

%======================================================
\subsection*{Voltage-tunable noise}
%======================================================

\begin{figure}[H]
  \centering
  \includegraphics[width=0.8\textwidth]{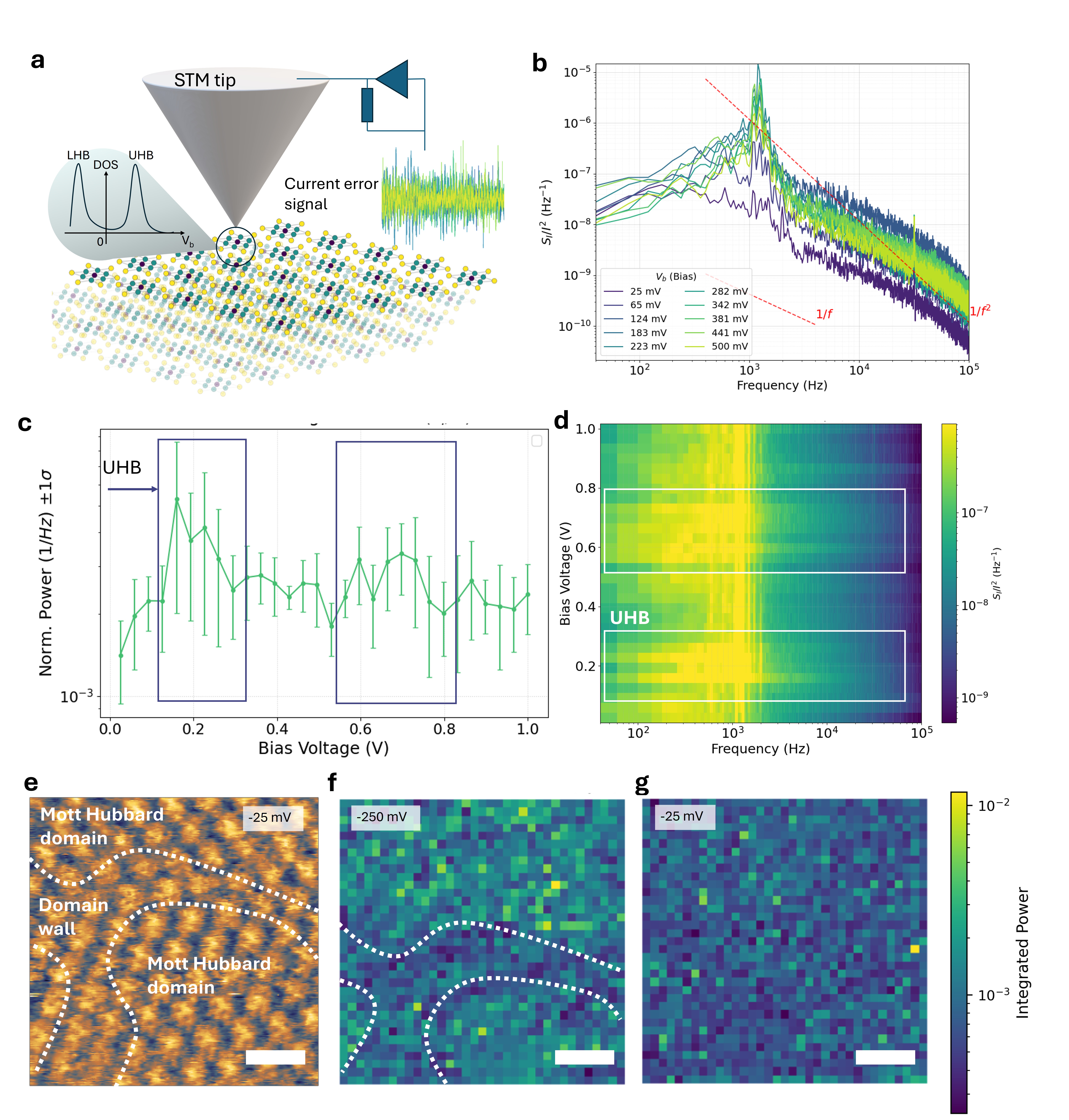}
  \caption{\textbf{Spatially resolved noise spectroscopy of the NC-CDW phase of 1T-TaS$_2$.} 
\textbf{(a)} Schematic illustration of the STM setup to measure the current error signal, the bias is applied to the tip. At \SI{300}{\kelvin} the UHB and LHB cannot be resolved with tunneling spectroscopy, but noise spectroscopy can detect them. 
\textbf{(b)} Evolution of the current PSD as a function of bias voltage obtained in the NC-CDW domain. The data has not been high-pass filtered (HPF). 
\textbf{(c)} Normalized integrated PSD versus bias voltage, after applying a \SI{2.5}{\kilo\hertz} HPF.  A pronounced maximum is observed at $\sim$200~mV, corresponding to the injection of electrons into the upper Hubbard Band (UHB). The peak between 600~mV and 800~mV corresponds to Ta 5d states. 
\textbf{(d)} 2D spectroscopic heatmap displaying the mean noise power as a function of frequency and bias voltage. The white boxes highlight the bias windows of enhanced noise at the UHB and the Ta 5d states. 
\textbf{(e)} STM topography ($V_{bias} = -25$~mV, $I_{set} = 1$~nA) of the NC-CDW surface. The dashed lines trace the Mott Hubbard domain walls (discommensurations) separating the domains. 
\textbf{(f, g)} Spatially resolved maps of the total integrated noise power at two tip biases.  At the UHB energy, the noise is spatially heterogeneous and inverted relative to conductivity: domains exhibit significantly higher noise power (bright) than the domain walls (dark). 
Inside the Mott gap, the noise power is effectively homogeneous and suppressed. Scale bar is equal to \SI{2}{\nano\meter}.}
  \label{fig:Noise_STM}
\end{figure}

First, we start by mapping current noise of the STM of a freshly cleaved 1T-TaS$_2$ crystal in UHV. We acquire noise spectra using STM by first imaging the NC-CDW with atomic resolution, as shown in \textbf{Figure \ref{fig:Noise_STM}a, e}, recorded at a sample bias of \SI{-25}{\milli\volt} applied to the tip i.e. tunneling in empty states of the sample. To limit thermal drift at \SI{300}{\kelvin}, we employ drift compensation during the measurements. Noise grid spectroscopy is obtained by measuring current traces (current error signal) at each grid point, in constant current mode. The power spectral density (PSD) \cite{Tamir2022ShotnoiseMicroscope} of a single grid point with varying tip bias voltages is given in \textbf{Figure \ref{fig:Noise_STM}b}. Note that the feedback introduces resonances and correlations up to around \SI{2.5}{\kilo\hertz} and we high pass filter the noise data. In \textbf{Supplementary \ref{sec:si_feedback}} we characterize the feedback loop. In the PSD, a pronounced peak is observed near \SI{20}{\kilo\hertz}, originating from mechanical resonances of the STM setup, and it does not depend on the bias voltage. These PSD spectroscopic data reveal a pronounced deviation from the $1/f^\alpha$ power law typically associated with collective sliding dynamics \cite{Liu2018LowFrequencyMaterials}.

Examining the bias dependent normalized PSD power in \textbf{Figure \ref{fig:Noise_STM}c}, we observe a clear enhancement in the bias range around \SI{200}{\milli\volt}. We associate this feature with electron injection into the upper Hubbard band (UHB), in the \textbf{Supplementary} we show for the filled state the similar structure. The energetic near-symmetry of this features about the Fermi level is the key discriminant: a Mott UHB is necessarily accompanied by an LHB at the mirrored bias, whereas an extrinsic depinning threshold (see below) has no reason to produce such a partner. These energies are close to the known Hubbard bands of 1T-TaS$_2$, which before have only been spatially resolved at low temperature in the commensurate (C) CDW phase \cite{Fei2022Understanding1T-TaSe2, Bu2019Possible1T-TaS2, Cho2017Correlated1T-TaS2}, owing to the high energy resolution of STS, which at \SI{300}{\kelvin} is degraded by thermal broadening \cite{Park2019EmergentWave}. To the best of our knowledge, this is the first atomic-scale detection of the Mott UHB signature at \SI{300}{\kelvin}, achieved through noise power rather than tunneling spectroscopy. An energy map of the noise power as a function of bias is displayed in \textbf{Figure \ref{fig:Noise_STM}d}. Similarly, it reveals a pronounced enhancement in PSD power in the range of \SI{600}{} to \SI{800}{\milli\volt}, which we associate with the Ta 5d conduction bands. We do want to emphasize that the Mott physics, and thus assigning the \SI{200}{\milli\volt} feature to the UHB, remains debated, in \textbf{Supplementary \ref{sec:MottDiscussion}} a further discussion is provided. 

Liu \textit{et al.} \cite{Liu2018LowFrequencyMaterials} investigated similar low-frequency current fluctuations in 1T-TaS$_2$ devices, interpreting observed Lorentzian spectral features as signatures of collective CDW depinning and sliding. They reported noise power maxima at the depinning threshold ($V_\text{DT} \approx 200$~mV) and corner frequencies shifting from $\sim$10~Hz to $\sim$80~kHz, extracting an activation energy of $E_a \approx 2.3$~eV. While this was attributed to the collective motion of 20--30~nm domains, our spatially-resolved grid noise spectroscopy, \textbf{Figure \ref{fig:Noise_STM}f, g} reveals a more complex picture that is incompatible with a purely collective sliding model alone, but our work is also applied to the surface of macro-scale crystals, where CDW pinning may be different than for exfoliated thin layers. In the spatial distribution of the noise: at the UHB energy ($\sim$200~mV), we observe that the domains exhibit significantly higher, by an order of magnitude, integrated noise power than the domain walls. Complementary evidence from our previous STM study of the NC-CDW state \cite{Verhage2026Interplay1T-TaS2} showed that (slowly) mobile discommensurations are present in the NC-CDW structure, evolving quasi-statically. While this confirms that collective CDW rearrangements do occur at room temperature, supporting the physical plausibility of the sliding mechanisms \cite{Lei2016StrongChains, Ghosh2025QuieterNanowires, LeBolloch2023TrackingDiffraction}, although these are mostly reported for (quasi) 1D CDW systems. Because we are still able to map the NC-CDW domain structure in the time of the grid noise measurement ($\sim$20min), our result point towards collectively slowly evolving yet highly localized CDW discommensurations reorganization, more akin to a local CDW ``jitter" instead of a pure collective sliding model. Furthermore, at the 25~mV bias where we directly image discommensuration motion before \cite{Verhage2026Interplay1T-TaS2}, our grid spectroscopy shows no obvious spatial contrast between domains and walls, indicating that the local electronic structure is also contributing to the noise power. 

We must also exclude extrinsic circuit instabilities as the source of these fast fluctuations. While Geremew et al. \cite{Geremew2020HighfrequencyConcept} showed that negative differential resistance (NDR) at the NC-incommensurate NC-(IC) phase transition can drive MHz-frequency oscillations that mimic sliding signatures ($f \propto I$), our measurements are performed at the UHB peak ($\sim$200~mV) in a regime of positive differential conductance ($dI/dV > 0$). The lack of NDR, together with the strong spatial confinement of the noise to NC-CDW domains, rules out a global circuit resonance as its origin. We therefore infer that the current fluctuations in 1T-TaS$_2$ stem from slow, collective CDW rearrangements instead of large hysteric NC-IC phase transitions \cite{Han2015ExplorationCrystallography, Yoshida2015MemristiveCrystals}, most likely linked to the glass-like NC-CDW structure \cite{Verhage2026Interplay1T-TaS2} and quenched disorder to which we will return later, which govern transport in thin flakes but evolve sluggishly at the surfaces of bulk crystals due to interlayer stacking coupling \cite{DeLaTorre2025DynamicQuench}. Using spatially resolved noise spectroscopy \cite{OrtegoLarrazabal2026CryogenicNoise-STM} can thus allow for mapping electronic features complementary to standard tunneling spectroscopy. This also offers pathways for future CDW dynamics mapping of similar materials.

%======================================================
\subsubsection*{Noise analysis}
%======================================================

\begin{figure}[h!]
    \centering
    \includegraphics[width=0.85\textwidth]{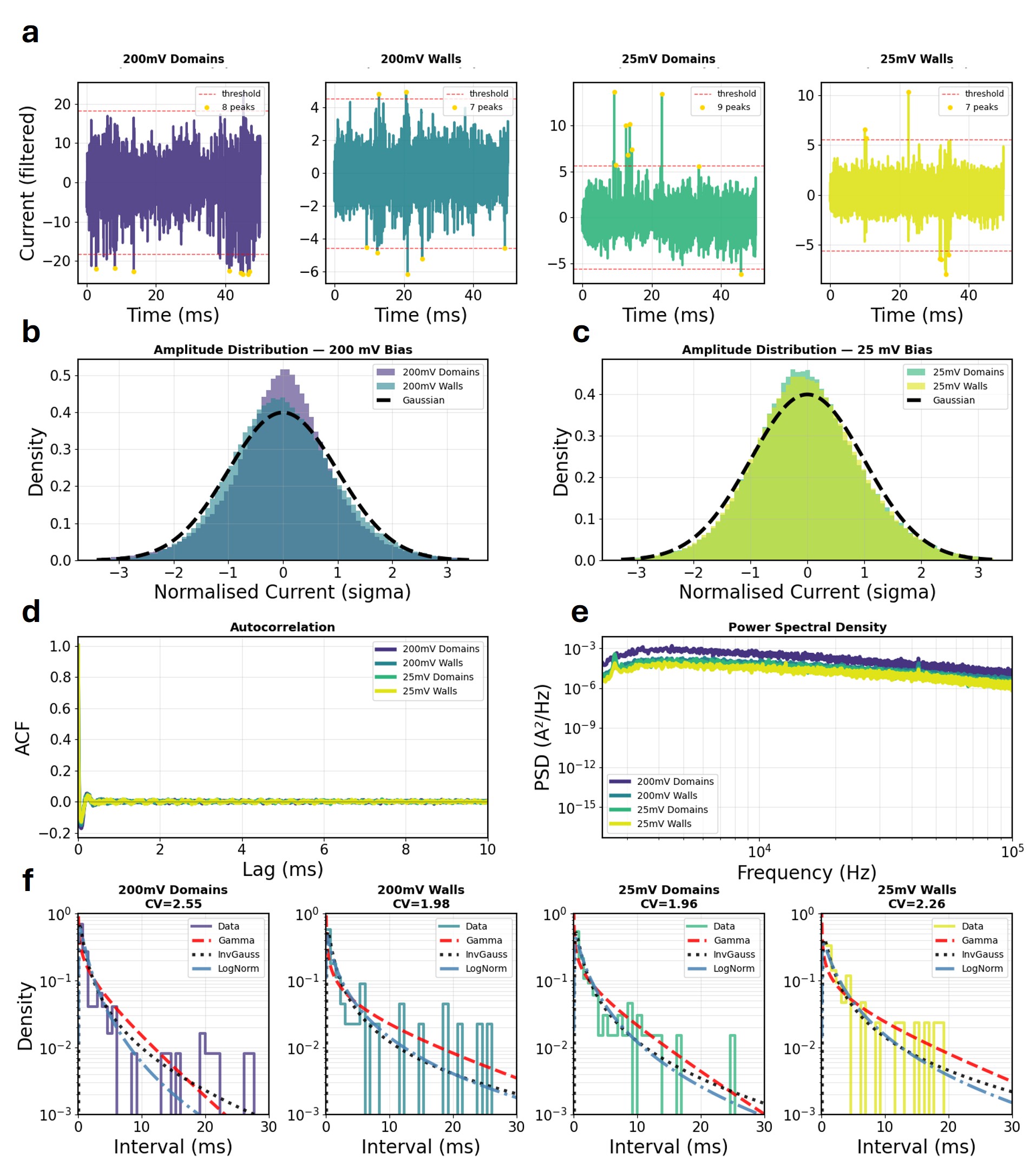}
    \caption{\textbf{Statistical noise analysis of $1T$-TaS$_2$ in the NC-CDW phase.} 
    \textbf{(a)} Normalized current time series for domains and wall regions at \SI{200}{\milli\volt} and \SI{25}{\milli\volt} bias, showing characteristic stochastic fluctuations and bursty current spikes.  
    \textbf{(b, c)} Amplitude distributions (PDFs) at \SI{200}{\milli\volt} and \SI{25}{\milli\volt}, respectively. Both regions exhibit leptokurtic behavior with heavy tails deviating from Gaussian overlays (dashed lines), particularly in the domains at \SI{200}{\milli\volt}.
    \textbf{(d)} Autocorrelation functions showing rapid decorrelation within \SI{1}{\milli\second}, indicating the absence of long-term periodic oscillations, after applying a \SI{2.5}{\kilo\hertz} HPF. 
    \textbf{(e)} PSD displaying $1/f^\gamma$ power-law roll-off.
    \textbf{(f)} Inter-Spike-Event (ISE) distributions on a semi-log scale fitted with Gamma, inverse Gaussian and LogNormal distributions. See \textbf{Supplementary \ref{sec:si_isi}} for all fits.}
    \label{fig:noise_analysis}
\end{figure}

In the following, we statistically analyze current vs time traces of the NC-CDW structure. The current contains discrete, burst-like fluctuations (spikes), which we treat as events, \textbf{Figure \ref{fig:noise_analysis}a}. The yellow markers in the plots highlight identified current bursts. We note these current features as possible Mott physics resulting in an emulator of the firing of biological neurons \cite{Stoliar2017LeakyIntegrateandFireInsulator, Pickett2013ScalableMemristors}. To map the characteristics of tunnel current bursts, we compared four conditions: in Mott-Hubbard domains and on domain walls, each measured with a low bias ($25\,\mathrm{mV}$, inside the Mott gap) and a bias coinciding with the upper Hubbard band (UHB, $\sim\!200\,\mathrm{mV}$). After removing the slow feedback loop effects with a zero-phase high-pass filter (\SI{2.5}{\kilo\hertz}), see \textbf{Supplementary \ref{sec:si_feedback}} for feedback loop characterization, burst events were detected per current traces (segments) with a threshold (median-absolute-deviation) of a factor 5, indicated with the red horizontal dashed line in \textbf{Figure \ref{fig:noise_analysis}a}. Three distinct, complementary aspects of the noise were quantified: (i) the size distribution of the current spikes, (ii) the distribution of time intervals between successive events, and (iii) whether those intervals carry temporal memory. The  quantities are summarized in \textbf{Table~\ref{tab:summary}} and \textbf{Figure \ref{fig:noise_analysis}}.
 
The simplest measure of spike size is the excess Kurtosis of the filtered current; how much heavier the tails of the amplitude distribution are than a Gaussian (for which the excess kurtosis is zero), \textbf{Figure \ref{fig:noise_analysis}b, c}. On Mott domains the distribution is strongly non-Gaussian (excess kurtosis $\approx 6.0$ at $200\,\mathrm{mV}$ and $5.6$ at
$25\,\mathrm{mV}$): large current spikes occur far more often than chance. On the metallic domain walls the same statistic is much smaller, and at the UHB bias it is closer to Gaussian with an excess kurtosis of $1.3$. The contrast is sharpest at the UHB, where the domains exceed the walls by roughly a factor of $4.5$.

The autocorrelation and power spectral density, \textbf{Figure \ref{fig:noise_analysis}d,e}, are calculated on the high-pass-filtered current and therefore characterize the broadband amplitude fluctuations. In all four conditions the autocorrelation decays within a fraction of a millisecond and is flat thereafter, with a shallow negative lobe at short lag that reflects the $2.5\,\mathrm{kHz}$ high-pass corner rather than a physical
anti-correlation; the filtered current is thus effectively short correlated on millisecond scales. Also after applying the HPF, the difference is the overall PSD level, with the $200\,\mathrm{mV}$ domains carrying the most power, in agreement with their larger and more frequent bursts and spatial contrast in \textbf{Figure \ref{fig:Noise_STM}}. 
 
We aimed to identify the functional form of the interval spike event (ISE); a direct emulator of neuronal firing behavior \cite{Liu2021TantalumProperties} by fitting its distribution using a maximum likelihood estimation approach with left-truncation at the detection refractory. Model comparison was performed using the Akaike Information Criterion \cite{Akaike1974NewIdentification}, as described in \textbf{Supplementary \ref{sec:si_isi}} and shown in \textbf{Figure \ref{fig:noise_analysis}f}. We fitted seven candidate distributions: exponential, gamma, inverse Gaussian, log-normal, Weibull, a two-timescale (biexponential) mixture, and a truncated power law \cite{Clauset2009PowerLawData}. Note that in the work of \cite{Liu2021TantalumProperties} the ISE was found to vary between gaussian, exponential and gamma depending on higher bias voltage values around \SI{2}{\volt}. In our work, we restricted to a lower bias voltage \SI{200}{\milli\volt} for tip stability. In our ISE, the gamma (single-timescale renewal) form is rejected in all conditions. Our ISE data do not discriminate what type of ISE distribution is dominant: the inverse-Gaussian, log-normal, Weibull, biexponential and truncated-power-law forms are statistically indistinguishable, with no model reaching an
Akaike weight above $0.51$. In the roughly one decade of intervals accessible here these heavy-tailed forms are mathematically near-degenerate. We therefore report the intervals as heavy-tailed and over-dispersed, especially for the \SI{200}{\milli\volt} domains, but we do not assign a specific distribution \cite{Liu2019EnergyEfficientComputing}. Likely, in the low bias regime the tunability between distributions is limited. The intervals between spike events are over-dispersed in every condition: their coefficient of variation (CV, the standard deviation divided by the mean) \cite{Shadlen1998VariableCoding} lies between $\approx 1.95$ and $2.55$, well above the value of $1$ expected for a memoryless (Poisson) process. This over-dispersion alone does not establish temporal clustering as a heavy-tailed renewal process is already over-dispersed, which is why we turn to the shuffle-controlled Fano test below.

Burstiness in the ISE histogram does not by itself imply temporal memory: a process with independent, heavy-tailed intervals already looks clustered. To separate genuine memory, important for LIF emulating neurons \cite{Kim2023SharingNetworks}, from this baseline we used a shuffle test \cite{Theiler1992TestingData}: randomly reordering the measured intervals preserves their distribution exactly while destroying any temporal correlation. The discriminating statistic is the Fano factor \cite{Nawrot2010AnalysisTrains, Farkhooi2009SerialVariability} (the variance-to-mean ratio of event counts in a fixed window). For a renewal process the Fano factor is fixed by the interval distribution alone and is therefore unchanged by shuffling; any residual drop measures serial correlation beyond the marginal distribution. We thus quantify burst memory as $\Delta\mathrm{Fano} =
\mathrm{Fano}_{\mathrm{original}} - \mathrm{Fano}_{\mathrm{shuffled}}$, evaluated over counting windows of two, five, and ten mean intervals and reported at the
five-interval scale ($\approx\!14\,\mathrm{ms}$ for the domains), with the shuffle distribution providing an error bar and a $z$-score. The $z$-score is the standardized drop, $z = \big(\mathrm{Fano}_{\mathrm{original}} -
\langle\mathrm{Fano}_{\mathrm{shuffled}}\rangle\big)/\sigma_{\mathrm{shuffled}}$, where $\langle\mathrm{Fano}_{\mathrm{shuffled}}\rangle$ and $\sigma_{\mathrm{shuffled}}$ are the mean and standard deviation of the Fano factor over the shuffled data. It expresses the measured clustering in units of the chance scatter expected from the finite event count, normalising away the differing event statistics of the four conditions; we take $z>3$ as the criterion for genuine sequence memory. Only one condition shows a significant effect: the $200\,\mathrm{mV}$ Mott domains, with $\Delta\mathrm{Fano} = +2.46$ ($z = +4.7$). The $200\,\mathrm{mV}$ walls show no drop ($\Delta\mathrm{Fano} = -0.36$, $z = -0.6$), and neither the $25\,\mathrm{mV}$ domains ($z = +0.5$) nor the $25\,\mathrm{mV}$ walls ($z = +1.3$) reach significance. Genuine sequence ISE memory is thus confined to the Mott domains at the UHB bias. To summarize, we identified large, heavy-tailed spikes on domains versus Gaussian-like noise on walls; bursty, heavy-tailed timing in all conditions; shuffle-robust memory only on the $200\,\mathrm{mV}$ domains; and UHB-peaked, domain-localized noise power. Note, we also measured the same noise segments on MoS$_2$ monolayer with the result given in \textbf{Supplementary \ref{sec:si_MoS2}}. For MoS$_2$ monolayer we observe a Gaussian currents segment behavior, as expected for a system without the complex CDW and its nonlinear response. 

\begin{table}[t]
\centering
\caption{Per-condition statistics of the tunnel current noise. Excess Kurtosis $K$ measures the heaviness of the
amplitude tails (0 = Gaussian); CV is the interval coefficient of variation (1 = Poisson) or dispersion;
$\Delta Fano{}$ is the shuffle-controlled Fano drop at the $\approx\!14\,\mathrm{ms}$ burst
window, with $z$ its significance. $N$ is the number of detected events (current spikes).}
\label{tab:summary}
\begin{tabular}{lrrrrr}
\toprule
Condition & $N$ & Excess Kurtosis $K$ & CV & $\Delta Fano{}$ & $z$ \\
\midrule
$200\,\mathrm{mV}$ domains & 169 & 5.99 & 2.55 & $+2.46$ & $+4.7$ \\
$200\,\mathrm{mV}$ walls   & 69  & 1.32 & 1.98 & $-0.36$ & $-0.6$ \\
$25\,\mathrm{mV}$ domains  & 97  & 5.55 & 1.96 & $+0.26$ & $+0.5$ \\
$25\,\mathrm{mV}$ walls    & 66  & 3.26 & 2.26 & $+0.64$ & $+1.3$ \\
\bottomrule
\end{tabular}
\end{table}

It is instructive to compare these spike trains with those of biological neurons. We state the comparison as an analogy at the level of spike statistics rather than a claim of true emulation. First, the heavy-tailed, over-dispersed intervals
(CV $\approx 2$--$2.5$) place the domain, but also the walls, neuron in the irregular, fluctuation-driven regime characteristic of cortical neurons \cite{Softky1993HighlyEPSPs, Shadlen1998VariableCoding}. The current spikes are thus consistent with a LIF neuron, though, as emphasized above, the noise at low bias voltage cannot distinguish this form from the other heavy-tailed distributions. The sign of the shuffle memory restricts which analogy is appropriate. The $200\,\mathrm{mV}$ domains exhibit a positive serial correlation ($\Delta\mathrm{Fano}>0$), matching the clustered firing of information-carrying neurons such as auditory-nerve fibers \cite{Lowen1992AuditorynerveScales}. The domain walls, by contrast, show no significant serial correlation ($z = -0.6$) and are renewal-like at the noise level; the regular-spiking analogy for the walls \cite{Farkhooi2009SerialVariability} emerges only after integration in the LIF circuit, not in the raw interval statistics.

%============================================================
\subsection*{Model}
%============================================================

\begin{figure}[b!]
    \centering
    \includegraphics[width=0.85\textwidth]{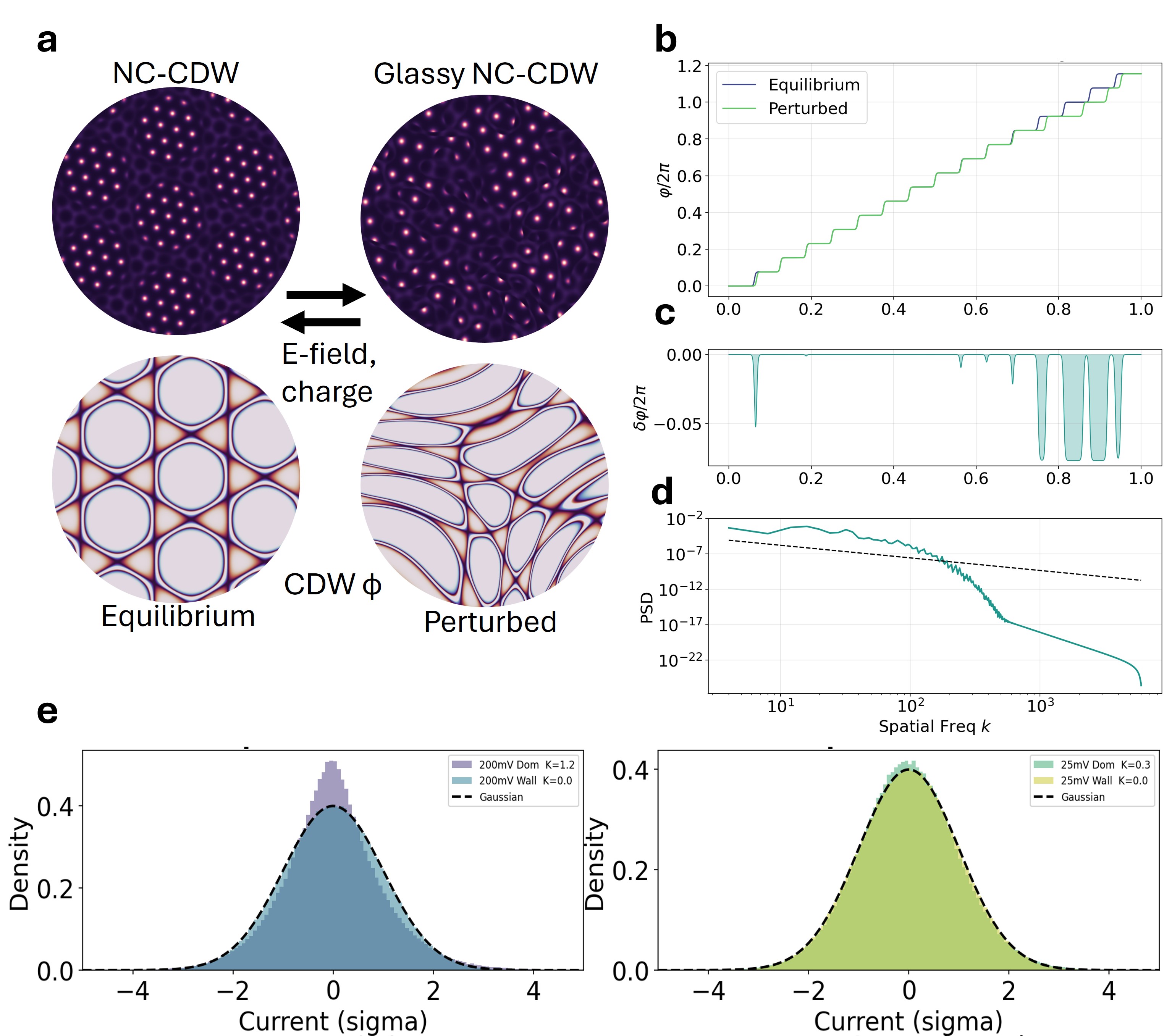}
    \caption{\textbf{Model of fractal discommensurations jamming as the origin of the non-linear current spike response.}
    \textbf{(a)}~Numerically computed NC-CDW (top) and phase (bottom); bright interiors are $\sqrt{13}\times\sqrt{13}$ commensurate domains, dark areas are discommensurations. Under STM tip electric field and/or charge injection, for example, the NC-CDW is disturbed into a correlated glass-like phase \cite{Verhage2026Interplay1T-TaS2}.
    \textbf{(b)}~Model CDW phase profile $\varphi(x)/2\pi$ comparing the
    equilibrium staircase (green), with the steps indicating the discommensurations and plateaus commensurate domains, with the fractal jammed state (blue) under perturbation. 
    \textbf{(c)} The displacement field $\delta\varphi$ is bursty
    and heavy-tailed, reflecting a cascade of fractal shifts of the phase discomensuration under perturbation.    
    \textbf{(d)}~CDW spatial frequency $k$ as function of the power spectral density, showing a fractal organization for large wavelengths and a roll-off due to thermalization, smoothening the discommensurations network. 
    \textbf{(e)}~Calculated amplitude distributions (in units of deviation, sigma) of the high-pass-filtered, simulated tunneling-current noise at \SI{200}{\milli\volt} and \SI{25}{\milli\volt} for domain-center and domain-wall positions. Leptokurtic behavior is reproduced for the \SI{200}{\milli\volt} domains. 
}
    \label{fig:Model_noise}
\end{figure}

We characterize the noise within a mesoscopic framework that connects the heavy-tailed, correlated bursts to avalanche-like dynamics of the discommensuration network. This approach views the CDW underneath the TM tip as undergoing a continuously evolving, self-organized reconstruction of CDW commensurability, accompanied by glass-like \cite{Verhage2026Interplay1T-TaS2} coherence dynamics \cite{Yilmaz2026Electronic2}. In \textbf{Figure \ref{fig:Model_noise}a}, the state of NC-CDW (amplitude) and its phase order are schematically shown to respond to external perturbations such as the electric field of the STM tip or charge injection \cite{Mraz2023ManipulationCrystal}. The dark purple spots indicate SoD (stars-of-david) structured in the (disordered) NC-CDW phase. In our earlier work \cite{Verhage2026Interplay1T-TaS2}, we revealed the fractal geometry of the 1D discommensuration network by imaging the CDW phase order parameter \cite{Pasztor2019HolographicParameter} with STM. Under quenched disorder or in the presence of structural defects, this discommensuration network assumes a fractal configuration with a Hausdorff dimension of about $D_f \sim 1.2$. We interpret the noise spectroscopy as probing the tunneling modulation (electronic contribution) generated by CDW structural rearrangements within a jammed, elastic discommensuration network that undergoes thermally and electric-field-assisted depinning (mechanical contribution). Strong depinning behavior previously reported in 1T-TaS$_2$ \cite{Mohammadzadeh2021RoomDevices} is, in our view, a signature of structural relaxation in this jammed, glass-like CDW state. Consequently, we adopt a Langevin-type description \cite{Verhage2026Interplay1T-TaS2}, in which the dynamics of the CDW phase discommensuration network $u_i(t)$ are governed by an overdamped elastic equation:

\begin{equation}
    \eta \frac{\partial u_i}{\partial t} =
    \left[ c_i (u_{i+1} - u_i) + c_{i-1} (u_{i-1} - u_i) \right]
    + k_0 (V_{drive}\, t - u_i) + D_{pin}\, \xi(u_i) + \zeta_i(t)
    \label{eq:langevin}
\end{equation}

In this framework, the respective terms describe the spatially non-uniform elastic coupling ($c_i$), the driving force due to the macroscopic bias ($k_0$), the commensurability pinning potential ($D_{pin}$), and a Langevin heat bath $\zeta_i(t)$ that captures thermal creep at $T = 300\,\text{K}$. Imposing heavy-tailed, spatially localized bond strengths $c_i$ maps the rigid hubs and weak bottlenecks of the 2D discommensurations fractal network onto the 1D description, such that the line preferentially yields at its weakly coupled bottlenecks and the resulting distribution of integrated avalanche sizes $S$ develops a heavy tail. Since each CDW phase slip event redistributes elastic stress to neighboring sites, these structurally driven slips are temporally correlated and occur in cascades, rather than as independent events. Compared to memoryless (Poisson), the observed phase-slip sequence is therefore over-dispersed ($CV > 1$), and not uniform or metronome-like. We further emphasize that, within this mesoscopic description, the over-dispersion and the serial burst memory quantified below both arise from the same underlying $S$ clustering, instead of representing independent control parameters. This mechanical CDW evolution in the tunnel juction is then converted into an electronic response via tunnel-current pulses: during structural CDW collapse in the tunnel junction, the Mott-gapped domains act as intense noise sources, whereas the metallic domain walls mainly contribute to low-frequency viscous creep.

\textbf{Figure \ref{fig:Model_noise}b} presents the structural phase profile $\varphi(x)$ as a spatial “staircase.” The flat plateaus correspond to individual CDW domains, while the vertical steps denote metallic discommensurations, i.e., domain walls \cite{McMillan1976TheoryTransition}. By superimposing the perturbed
avalanche configuration (blue curve) on the equilibrium profile (green curve), this figure reveals that the domain walls do not wander independently; instead, they move in a correlated, hierarchical fashion that reflects the collective elastic response of the CDW. To isolate the transient aspects of this structural evolution, \textbf{Figure \ref{fig:Model_noise}c} displays the spatial displacement field, $\delta\varphi(x)$. The phase modulations (peaks in the trace) directly encode the phase slips that produce the tunneling current fluctuations, since domains and domain walls host distinct electronic states \cite{Park2019EmergentWave}. The sharp, localized spikes correspond to abrupt structural phase shifts in which the accumulated elastic stress overcomes the local pinning potential. We associate these CDW phase avalanche events with the origin of the heavy-tailed current bursts. In addition, the continuous, low-amplitude “jitter” along the baseline reflects the $300\,\text{K}$ thermal creep, signifying that the CDW phase continues to fluctuate within its quenched disordered potential even in the intervals between large avalanches. In \textbf{Figure \ref{fig:Model_noise}d}, we evaluate the spatial PSD of the displacement field to probe the fractal organization of the glassy CDW superlattice. At low spatial frequencies ($k < 100$), where $k$ is the inverse of the CDW spatial period, the PSD exhibits a $k^{-1.78}$ power-law decay, indicating that the long-range domain wall network assumes a self-affine fractal geometry. This exponent quantifies the
self-affine roughness of the 1D displacement field in the model ($S(k) \propto k^{-(1+2\zeta)}$, implying a roughness exponent $\zeta \approx 0.39$). At higher spatial frequencies, the spectrum bends over,
signaling the thermal crossover where the $300\,\text{K}$ Langevin heat bath dominates and smooths out the finer nanoscale discommensuration features.

We consider the case in which an NC-CDW phase slip propagating through a commensurate domain, under the STM tip, introduces a local perturbation that distorts the underlying SoD superlattice \cite{Dong2023Emergent1T-TaS$_2$}. Rather than a uniform Mott-gap collapse \cite{Bu2019Possible1T-TaS2}, this transient strain drives a
localized reconfiguration of the SoD, triggering cascading ``melting'' and ``recrystallization'' of SoD clusters. The relevant electronic states are the UHB states. Because these states are sharp and energetically localized, a structural avalanche transiently displaces the UHB peak within the tunneling window set by the bias voltage, modulating the spectral weight available for tunneling.

In this transducer framework, the burst source does not need to be specific to the UHB: the structural avalanche perturbs the LDOS at essentially any bias, in line with the heavy-tailed statistics we detect on the domains even at the in-gap bias of \SI{25}{\milli\volt} (excess kurtosis $\approx 5.6$). What is unique to the UHB is the resulting electronic amplification and the emergence of temporal memory. When the sharp UHB peak falls within the bias window, tunneling into the singly occupied cluster activates a doublon \cite{Mann2016ProbingSpectroscopy}-recombination channel. The rate $\lambda$ of clustered recombination aftershocks initiated per primary structural avalanche is set by the bias overlap integral:

\begin{equation}
    \lambda = \mu \int \Delta\rho_{\text{UHB}}(E)
    \left[ f(E - eV_{\text{bias}}) - f(E) \right] dE
    \label{eq:doublon_rate}
\end{equation}

Here, $\mu$ denotes how many phase after shocks each primary avalanche triggers, $\Delta\rho_{\text{UHB}}(E)$ is the bias-induced modulation of the UHB density of states, and $f(E)$ is the Fermi–Dirac distribution. At \SI{25}{\milli\volt}, this integral is nearly zero, so the observed dynamics are purely structural. At \SI{200}{\milli\volt}, however, this active channel controls both the integrated noise power (\textbf{Figure \ref{fig:Noise_STM}f}) and the serial burst memory (\textbf{Figure \ref{fig:noise_analysis}}), which is markedly positive only in the \SI{200}{\milli\volt} domains ($z \approx 4.7$). Our focus is on reproducing the sign and bias dependence of this memory, rather than matching its exact magnitude: the doublon channel contributes an extra bias-gated serial correlation at \SI{200}{\milli\volt}, superimposed on the intrinsic structural clustering. This makes the two domain biases distinguishable in terms of memory, despite sharing the same underlying avalanche train. The aftershock process serves as a mesoscopic stand-in for doublon recombination \cite{Ligges2018Ultrafast2}. When an electron is injected into the UHB, it forms a highly localized, excited doublon state. This localized energy burst functions as a secondary trigger, launching a temporally clustered sequence of smaller structural CDW phase slips.

In our framework, the evaluation proceeds as follows. We first take the spatial integral of the Langevin equation, obtaining a global phase-slip velocity. Structural avalanches are then identified by applying a threshold to this velocity, producing a sequence of discrete phase-slip events specified by their onset times and spatially integrated sizes ($S$). Next, we incorporate the electronic response through a nonlinear transfer function, $\Phi(S) = S_{\text{sat}}(1 - \exp(-S/S_{\text{sat}}))$, which imposes an upper bound on the spectral weight transfer. This captures the discrete changes in local fractional filling and the corresponding modulation of the LDOS beneath the STM tip. The resulting transferred spectral weight is integrated over the STM measurement window set by the bias voltage and the Fermi-Dirac distribution, thereby mapping each CDW phase-slip size onto a discrete tunneling current burst amplitude.

The resulting synthetic time series is processed with a second-order Butterworth high-pass filter ($f_c = 2.5\,\text{kHz}$), mirroring the treatment of the experimental data. Noise statistics are directly computed from this filtered signal, and the resulting current distributions are presented in \textbf{Figure \ref{fig:Model_noise}e}. For simulated measurements on a CDW domain, the large, sparse current spikes associated with rapid shifts of the localized UHB peak through the tunneling window propagate through the high-pass filter and generate heavy-tailed amplitude distributions (excess kurtosis $\approx 5$--$6$ on the domains). By contrast, at metallic domain walls the transfer function saturates within the low-bias window, while the broad metallic background increases with $V_{\text{bias}}$: at \SI{200}{\milli\volt}, the wall pulses are overwhelmed by this background and the filtered distribution becomes nearly Gaussian (excess kurtosis $\approx 1.3$), whereas in the narrower \SI{25}{\milli\volt} window they partially reappear (excess kurtosis $\approx 3.3$). In both regimes, the walls mainly contribute low-frequency viscous creep, which is largely removed by the \SI{2.5}{\kilo\hertz} high-pass filter, leaving them effectively free of the serial burst memory characteristic of the domains.

%======================================================
\subsection*{LIF spike generator}
%======================================================

%~~~~~~~~~~~~~~~~~~~~~~~~~~~~~~~~~~~~~~~~~~~~~~~~~~~~
\begin{figure}[b!]
\centering
\includegraphics[width=\linewidth]{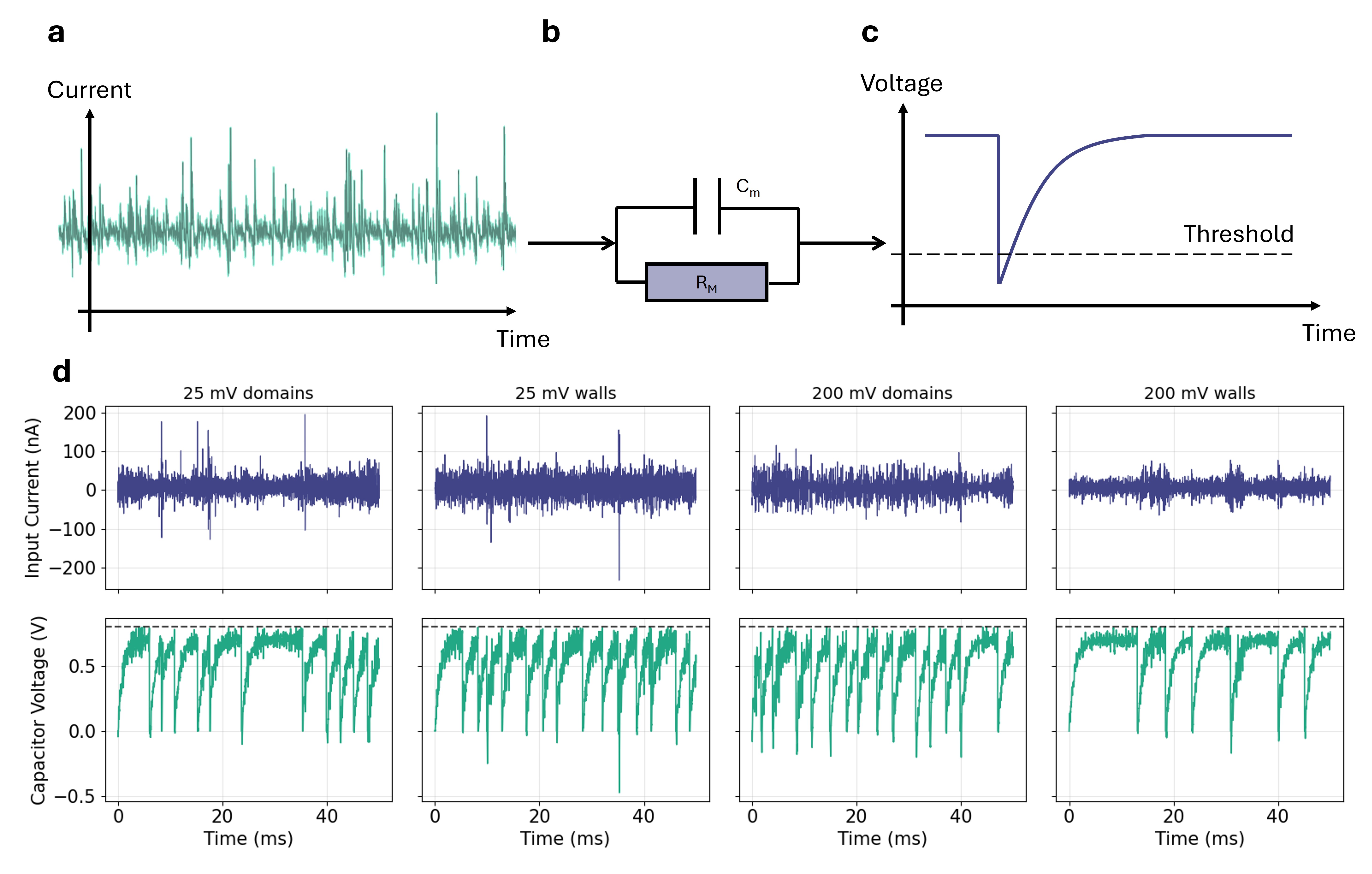}
\caption{\textbf{Emulation of a LIF circuit driven by measured  tunnel current noise.} 
(\textbf{a-c}) Conceptual schematic of the hardware emulation using real noise. The continuous, heavy-tailed current noise \textbf{(a)} is injected into an $RC$ integrator (\textbf{b}) representing the LIF neuron, utilizing a parallel capacitor ($C_m$) and leakage resistor ($R_M$). (\textbf{c}) The circuit integrates the charge bursts, firing a discrete action potential upon crossing a threshold before resetting. 
(\textbf{d}) Emulated hardware dynamics comparing the input current (top row) and resulting capacitor voltage (bottom row) across four distinct spatial (domains vs. walls) and bias (\SI{200}{\milli\volt} and \SI{25}{\milli\volt}) conditions.}
\label{fig:LIF_Simulator_spikes}
\end{figure}
%~~~~~~~~~~~~~~~~~~~~~~~~~~~~~~~~~~~~~~~~~~~~~~~~~~~~

In the last part we turn to assess whether 1T-TaS$_2$'s complex CDW fluctuations can serve as an intrinsic driver for neuromorphic primitives such as spiking oscillators \cite{Zhu2018LightTunableOscillators}. We did not fabricate a device but opted to simulate a sub-threshold LIF neuron with injecting real measured current bursts of the domains and walls at \SI{200}{\milli\volt} and \SI{25}{\milli\volt}; related behavior has been studied previously at the device level \cite{Khitun2017TwoDimensionalDevices, Khitun2018TransistorLessDevices}. To place the four conditions on an equal footing and remove the trivial dependence on absolute current magnitude, each high-pass-filtered trace was set to zero mean and unit RMS before injection; the dimensionless gain $g$ therefore sets the noise intensity in units of each trace's own RMS, so that comparisons across conditions probe the shape of the amplitude distribution and its temporal ordering at matched variance, not the physical current scale.

To test these primitives in a hardware-realistic setting we implemented an analog RC emulation (\textbf{Figure \ref{fig:LIF_Simulator_spikes}a-c}): a \SI{10}{\pico\farad} membrane capacitor $C_m$ in parallel with a \SI{100}{\mega\ohm} leak resistor $R_M$, giving a membrane time constant $\tau_m = R_M C_m = \SI{1}{\milli\second}$. The high-pass-filtered currents ($>\SI{2.5}{\kilo\hertz}$) were scaled to an RMS amplitude of \SI{20}{\nano\ampere} and injected with a \SI{7}{\nano\ampere} DC bias, holding a steady-state baseline of \SI{0.7}{\volt}, \SI{0.1}{\volt} below the \SI{0.8}{\volt} firing threshold. 

\begin{figure}[h!]
\centering
\includegraphics[width=\linewidth]{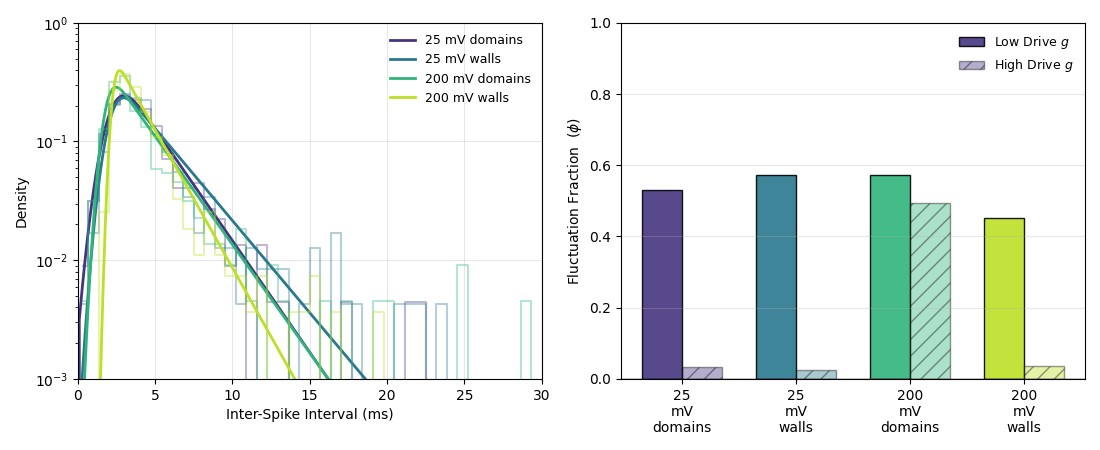}
\caption{\textbf{Statistical decomposition of LIF inter-spike intervals.} 
(\textbf{a}) ISE distributions for the four spatial and bias conditions of the numerical LIF spikes. The ISE is best fitted by an ex-Gaussian maximum likelihood fit (solid lines). This model  maps the ISE into a deterministic integration drift time ($\mu$) and a stochastic fluctuation wait time ($\tau$). 
(\textbf{b}) The fluctuation fraction, defined as $\phi = \tau / (\mu + \tau)$, comparing the low-drive $g$ baseline (solid bars) to the high-drive $g$ regime (hatched bars). Low $g$ forces all conditions to act as stochastic burst-detectors at low drive ($\phi \approx 0.45\text{--}0.60$). Between $g$ of 1.5 and 4, \SI{25}{\milli\volt} conditions and \SI{200}{\milli\volt} walls regularize into coherent spiking ($\phi \to 0$), whereas the \SI{200}{\milli\volt} domains maintain a fluctuation fraction ($\phi \approx 0.49$) with non-regular bursting.}
\label{fig:LIF_ISE_gain}
\end{figure}

Driven by the NC-CDW tunneling noise at \SI{20}{\nano\ampere} RMS, the emulator remained stable across all four conditions (\textbf{Figure \ref{fig:LIF_Simulator_spikes}d}). At fixed variance, a leptokurtic input concentrates most of its sample mass near zero and carries its variance in rare bursts; the \SI{200}{\milli\volt} domains therefore hold the capacitor quiet between events and fire clean, isolated action potentials only when a large phase slip arrives, consuming minimal dynamic power in the intervals. The near-Gaussian walls behave oppositely: their fluctuations are spread across many moderate events, so once the drive is sufficient the membrane integrates toward threshold in a regular, drift-dominated manner rather than waiting on rare bursts, the clock-like firing recovered quantitatively in \textbf{Figure \ref{fig:LIF_ISE_gain}} ($\phi \to 0$). These results demonstrate the tunability of the LIF neuron by bias and spatial position at the atomic scale.

%~~~~~~~~~~~~~~~~~~~~~~~~~~~~~~~~~~~~~~~~~~~~~~~~~~~~
%~~~~~~~~~~~~~~~~~~~~~~~~~~~~~~~~~~~~~~~~~~~~~~~~~~~~

Finally, we analyzed the ISE distributions of the emulated LIF firing output. As discussed before, the current noise bursts itself did not yield a conclusive distribution of the ISE. However, applying the same distributions fits to the real noise injected into the stimulated LIF circuit does yield a dominant distribution: for all four spatial and bias conditions, the ISE distribution is dominated by an ex-Gaussian profile (Akaike weight $\approx 1.0$). This ex-Gaussian model mirrors the dual mechanisms of the LIF membrane: a Gaussian component captures the regular, deterministic drift as the membrane integrates baseline current to threshold, while an exponential tail captures the stochastic waiting time for a large noise fluctuation to force a spike. In this ex-Gaussian decomposition, the inter-spike interval is split into two competing physical mechanisms. The Gaussian mean ($\mu$) captures the deterministic drift or the predictable time required for the membrane to integrate the baseline current toward the firing threshold. The exponential tail ($\tau$) represents the stochastic fluctuation, the waiting time required for a large noise burst to finally push the voltage over the edge. The balance between these two regimes we quantify by the fluctuation fraction, $\phi = \tau / (\mu + \tau)$. When $\phi \to 0$, the deterministic drift dominates, yielding highly regular, metronome-like firing. As $\phi \to 1$, the stochastic wait time dominates, resulting in highly irregular, burst-driven firing governed entirely by sudden atomic phase slips. While the 25 mV domains, 25 mV walls, and 200 mV walls all exhibit a transition toward clock-like deterministic firing at high drive $g$ (low $\phi$) the 200 mV domains resist this regularization for intermediate $g$. The 200 mV domains maintain a fluctuation fraction and high over-dispersion. This demonstrates that a single 1T-TaS$_2$ crystal contains two distinct, \textit{in situ}-tunable neuromorphic primitives: a coherent regular LIF firing (walls) and a burst-driven source (\SI{200}{\milli\volt} domains).

\section{Conclusion}

Using STM, we show that atomic-scale tunnel-current noise in the NC-CDW phase of 1T-TaS$_2$ is a voltage- and spatially-dependent statistical signal. The current fluctuations over the Mott domains are strongly heavy-tailed at both biases (excess kurtosis $\approx 6$),  whereas the metallic domain walls are less heavy-tailed and become near-Gaussian at the upper Hubbard band (excess kurtosis $\approx 1.3$). All four conditions are over-dispersed (CV $\approx 2$), so none is memoryless in the Poisson sense; what distinguishes the UHB-biased domains is genuine serial burst memory ($\Delta\mathrm{Fano} = +2.46$, $z = 4.7$), absent from the walls and from the in-gap domains. A mesoscopic model reproduces this four-condition behavior by converting fractal phase avalanches of the discommensuration network into current pulses, with a bias-gated doublon-recombination channel supplying the extra correlation unique to the UHB. Consequently, we interpret the room-temperature memristive behavior not as a bulk metal-to-band-insulator transition but as a continuous, fractal-organized deformation of the CDW discommensurations, under external perturbation such as electric field of the STM tip. A numerical leaky integrate-and-fire emulation shows that this tunneling noise produces two complementary, electrically and spatially selectable primitives in one material: a stochastic, burst-correlated domain channel and a near-Gaussian wall channel that emits more regular firing, both selectable by bias and tip position. Our work indicates that correlated fluctuations on the atomic scale of 1T-TaS$_2$ crystals can function as neuromorphic emulators by switching between metastable CDW states, with potential application in neuromorphic devices. Lastly, tunneling noise spectroscopy can be a valuable tool to probe the rich dynamics of CDW. 

\section{Methods}

\subsection*{Scanning tunneling and noise spectroscopy}
Experiments were performed on cleaved 1T-TaS$_2$, 2D Semiconductors, flux grown. Crystals were cleaved in UHV using the scotch tape metod. STM was performed using a Scienta Omicron VT-SPM operating in UHV, \SI{1e-9}{\milli\bar}. Bias voltage was applied to the STM tip, made of mechanically cut PtIr wire. Current-time measurements were recorded at fixed tip locations while stepping the bias voltage, using the nOmicron \cite{OGordon1002019OGordon100Conditioning} module for Python communication with a Scienta Omicron Matrix controller. A SRS SR830 DSP lock-in was used to calibrate the feedback loop. 

\section*{Acknowledgments}

Funding is provided by Eindhoven University of Technology.

\subsubsection*{Competing interests:}
There are no competing interests to declare.

\newpage

\bibliographystyle{unsrt} 
\bibliography{references}

\end{document}